\documentclass[prl,twocolumn,superscriptaddress,showpacs,floatfix]{revtex4}
\usepackage{graphicx,amsmath,amssymb,bm}

\usepackage{braket}
\usepackage{amsfonts}

\newcommand*{\intsum}{\mathop{\ooalign{\raise0.2pt\hbox{$\int$}%
            \cr\lower0.3pt\hbox{$\Sigma$}}}}

\begin{document}

\title{Quenching of spectroscopic factors for proton removal in oxygen isotopes}

\author{\O.~Jensen}
\affiliation{Department of Physics and Center of Mathematics for Applications, University of Oslo, N-0316 Oslo, Norway}
\author{G.~Hagen}
\affiliation{Physics Division, Oak Ridge National Laboratory,
Oak Ridge, TN 37831, USA}
\affiliation{Department of Physics and Astronomy, University of
Tennessee, Knoxville, TN 37996, USA}
\author{M.~Hjorth-Jensen}
\affiliation{Department of Physics and Center of Mathematics for Applications, University of Oslo, N-0316 Oslo, Norway}
\author{B.~Alex Brown}
\affiliation{National Superconducting Cyclotron Laboratory, Michigan
  State University, East Lansing, MI 48824-1321, USA}
\affiliation{Department of Physics and Astronomy, Michigan State University, East Lansing, MI 48824, USA}
\author{A.~Gade}
\affiliation{National Superconducting Cyclotron Laboratory, Michigan
  State University, East Lansing, MI 48824-1321, USA}
\affiliation{Department of Physics and Astronomy, Michigan State University, East Lansing, MI 48824, USA}

\begin{abstract}
  We present microscopic coupled-cluster calculations of the
  spectroscopic factors for proton removal from the closed-shell 
  oxygen isotopes $^{14,16,22,24,28}$O with the chiral nucleon-nucleon 
  interaction at next-to-next-to-next-to-leading order.
  We include coupling-to-continuum degrees of freedom by using a 
  Hartree-Fock basis built from a Woods-Saxon single-particle
  basis. This basis treats bound and continuum states on an equal footing.
  We find a significant quenching of spectroscopic factors in 
  the neutron-rich oxygen isotopes, pointing to enhanced many-body 
  correlations induced by strong coupling to the scattering continuum
  above the neutron emission thresholds. 
\end{abstract}


\maketitle

The concept of independent  particle motion, and 
mean-field approaches based thereupon, has played
and continues to play  a fundamental role in studies of quantum mechanical 
many-particle systems. From a theoretical standpoint, 
a single-particle (or quasiparticle) picture of states near the Fermi surface 
offers a good starting point for studies of systems with many
interacting particles. For example, the success of the nuclear shell
model rests on the assumption that the wave functions used in nuclear
structure studies can be approximated by Slater determinants built on
various single-particle states.  
The nuclear shell model assumes thus that protons and neutrons move as
independent particles  
with given quantum numbers, subject to a  mean field generated by all other nucleons. 
Deviations from such a  picture have been interpreted as a possible
measure of correlations. Indeed, 
correlations are expected to reveal important features of both the
structure and the dynamics of a many-particle system beyond the
mean-field picture.  

In a field like nuclear physics, where the average density in nuclei is high 
and the interaction between nucleons is strong, correlations 
beyond the independent-particle motion are expected to play an important role
in spectroscopic observables. In particular, experimental programs in
low-energy nuclear physics aim at extracting  
information at the limits of stability of nuclear matter. Correlations
which arise when moving towards either the proton or the neutron
dripline should then provide us with a better understanding 
of shell structure and single-particle properties of nuclei.
So-called magic nuclei are particularly important for a fundamental  
understanding of single-particle states outside shell closures, with
wide-ranging consequences spanning from our basic understanding of
nuclear structure to the synthesis of the elements
\cite{vijay1997,katenature}. 
Unfortunately, the correlations in many-particle
systems are  very difficult to quantify experimentally and to interpret
theoretically. There are rather few observables from which clear
information on correlations beyond an independent 
particle motion in a nuclear many-body environment can be extracted.    

A quantity which offers the possibility to study deviations from a
single-particle picture, and thereby provide information on
correlations, is the spectroscopic factor (SF).  
From a theoretical point of view they quantify what fraction
of the full wave function can be interpreted as an independent single-particle or 
hole state on top of a correlated state, normally chosen to be a
closed-shell nucleus.  
Although not being an experimental 
observable {\em per se} \cite{dick2002,akram2010,jennings2011}, 
the radial overlap functions, whose norm are the SFs, 
are required inputs to the theoretical models
for nucleon capture, decay, transfer and knockout reactions.
There is a wealth of experimental data and theoretical analysis of 
such reactions for stable nuclei 
\cite{vijay1997,carlo2009,jenny2009}. 
Data from $(e,e'p)$ experiments on stable nuclei 
\cite{vijay1997} indicate that proton absolute SFs are quenched
considerably with respect to 
the independent particle model value, with short-range and tensor
correlations assumed to be the main mechanism. Adding long-range
correlations as well from excitations around the Fermi surface, one
arrives at a quenching of $30-40\%$, see for example
Ref.~\cite{wimcarlo2004}. Nuclear physics offers 
therefore a unique possibility, via studies of quantities like SFs, 
to extract information about correlations
beyond mean-field in complicated, two-component, many-particle systems.

Recent data on knockout reactions on nuclei with large neutron-proton
asymmetries indicate that the nucleons of the deficient species, being
more bound, show larger reductions of spectroscopic strength than the
more weakly bound excess species
\cite{alexandra2008,alexandra2004}. It is the aim of this work to
understand which correlations are 
important when one moves towards more weakly bound systems. For this,
we study the chain of oxygen isotopes and compute SFs for proton
removal from  $^{14,16,22,24,28}$O. These isotopes 
span a large range of proton-neutron asymmetries, from $4/3$ in
$^{14}$O to $2/5$ in $^{28}$O. Using {\em ab initio} coupled-cluster
theory described below \cite{bartlett}, 
we argue that the reduction in SFs is
due to many-body correlations arising from the coupling to the
scattering continuum in neutron-rich  oxygen isotopes. After these
introductory remarks, we give a brief overview of our formalism,
before presenting our results and conclusions.  

The spectroscopic factor  
$S_{A-1}^A(lj) = \left|O_{A-1}^A(lj;r)\right|^2$,
is the norm of the overlap function,
\begin{equation}
O_{A-1}^A(lj;r) = \intsum_n\Braket{A-1||\tilde{a}_{nlj}||A}\phi_{nlj}(r).
\label{equDefOverlap}
\end{equation}
Here, $O_{A-1}^A(lj;r)$ is the radial overlap function of the many-body
wave functions for the two independent systems with $A$ and $A-1$
particles, respectively.
$\ket{A}$ and $\ket{A-1}$ can in general either be in their ground- or 
any excited state. In this work we consider only
overlaps with $\ket{A}$ and $\ket{A-1}$ in their ground states. 
The double bar denotes a reduced matrix
element, and the integral-sum over $n$ represents both the sum over the
discrete spectrum and an integral over the corresponding continuum
part of the spectrum. 
The annihilation operator $\tilde{a}_{nlj}$
is a spherical tensor of rank $j$. 
The radial single-particle basis function is given by the term
$\phi_{nlj}(r)$, where $l$ and $j$ denote the single-particle orbital and
angular momentum, respectively, and $n$ is the nodal quantum number. The isospin
quantum number has been suppressed.  We emphasize
that the overlap function, and hence also its norm, is defined
microscopically and independently of the single-particle basis. It is
uniquely determined by the many-body wave functions $\ket{A}$ and
$\ket{A-1}$. From the definition of the overlap function 
in Eq.~(\ref{equDefOverlap})
it is clear that the SF is mainly a measure of how well 
nucleus $A$ can be described by a single, uncorrelated nucleon
attached to nucleus $A-1$. 
Large deviations from unity indicate an increased role of
many-body correlations beyond a mean-field picture. For calculational
details see Ref.~\cite{Jensen2010}. 

We use the coupled-cluster (CC) ansatz \cite{bartlett} 
$\ket{\psi_0} = \exp{(T)}\ket{\phi_0}$
for the ground states of the closed-shell oxygen isotopes $^{14,16,22,24,28}$O.
The reference state, $\ket{\phi_0}$, is an antisymmetric product state
for all $A$ nucleons.  The cluster operator $T$ introduces correlations
as a linear combination of particle-hole excitations $T = T_1 + T_2 +
\ldots + T_A$, where $T_n$ represents an $n$-particle-$n$-hole
excitation operator.  For the CC singles and doubles
approximation (CCSD) employed in this work, $T$ is truncated at the
level of double excitations, $T = T_1 + T_2$. 

Due to the non-hermiticity of the standard CC formalism, 
we need to calculate both the left and the right eigenvectors. These are determined via
the equation-of-motion CC (EOM-CC) approach as
$\ket{A} \approx \ket{R_\nu^{A}(J_{A})} \equiv \exp{(T)} R_\nu^{A}(J_{A}) \ket{\phi_0}$ and
$\bra{A} \approx \bra{L_\nu^{A}(J_{A})} \equiv\bra{\phi_0}
L_{\nu}^{A}(J_A) \exp{(-T)} \label{equLA}$. The operators 
$R_\nu^A(J_A)$ and $L_\nu^A(J_A)$ produce linear
combinations of particle-hole excited states when acting to
the right and left, respectively. In the spherical form of the EOM-CC approach, the
operators have well defined angular momentum by construction, 
as indicated by $J_A$, which stands for the angular
momentum considered. If the $A$-body system is in its
ground state, the right EOM-CC wave function is identical to the
CC ground state.

Solutions for the $A-1$-body systems are
obtained with particle-removed equation-of-motion 
coupled-cluster method, 
where we use the CCSD ground state solution of the closed-shell
nucleus $A$ as the reference state in order to determine the 
corresponding left and right eigenvectors
$\ket{A-1} \approx\ket{R_\mu^{A-1}(J_{A-1})} \equiv \exp{(T)} R_\mu^{A-1}(J_{A-1}) \ket{\phi_0}$
and $\bra{A-1} \approx \bra{L_\mu^{A-1}(J_{A-1})} \equiv\bra{\phi_0} L_\mu^{A-1}(J_{A-1}) \exp{(-T)}$.
In actual calculations, the EOM-CC wave functions are obtained by determining
the operators $R_\nu^A(J_A)$ and $L_\nu^A(J_A)$ as eigenvectors of the
similarity-transformed Hamiltonian, $\overline H=\exp{(-T)}H\exp{(T)}$.
We refer the reader to
Refs.~\cite{bartlett, Hagen2010, Hagen2010a} for details about EOM-CC.

Finally, we can approximate the SF 
in the spherical CC formalism as
\begin{align}
S_{A-1}^{A}(lj) &= \intsum_n
	\braket{L^{A-1}_\mu(J_{A-1})|| \overline{\tilde a_{nlj}} || R^A_\nu(J_A)}
	\nonumber\\\times
	&\braket{R^{A-1}_\mu(J_{A-1})|| \overline{\tilde a_{nlj}} || L^A_\nu(J_A)}^*,
	\label{equSFdefinition}
\end{align}
where we have used the similarity-transformed spherical annihilation operator
defined in Ref.~\cite{Jensen2010}.
The labels $\mu$ and $\nu$ are included to distinguish between states
in $\ket{A}$ and $\ket{A-1}$.

The  intrinsic $A$-nucleon Hamiltonian reads
 $\hat{H} = \hat{T}-\hat{T}_{\rm cm} +\hat{V}$,
where $\hat T$ is the kinetic energy, $\hat T_{\rm cm}$ is the
kinetic energy of the center-of-mass coordinate, and $\hat V$ is the two-body
nucleon-nucleon (NN) interaction. We employ here the N$^3$LO model of Entem and
Machleidt \cite{Entem2003}. This interaction model is constructed with
a cutoff of $\Lambda=500$ MeV. 
CC calculations starting
from this Hamiltonian have been shown to generate solutions
that are separable into a Gaussian center-of-mass wave function and
an intrinsic wave function, see for example
Refs.~\cite{Hagen2010a,HPD09}.

We use a Hartree-Fock (HF) solution for the reference state, 
as detailed in for example Ref.~\cite{Hagen2010}. These HF 
solutions were built from the standard harmonic oscillator (HO) basis 
combined with Woods-Saxon (WS) single-particle bound- and scattering
states for selected partial waves. The role of the continuum is expected to be
important close to the dripline, as seen in Refs.~\cite{Hagen2010,michel2009,Volya2006}.
For this purpose we use a spherical WS basis 
for the neutron $s_{1/2}$, $d_{3/2}$, and $d_{5/2}$ partial
waves. The single-particle bound and scattering states are 
obtained by diagonalizing a one-body Hamiltonian with a 
spherical Woods-Saxon potential defined on a
discretized set of real momenta. We employ a
total of 30 mesh points along the real momentum axis 
for each of the $s_{1/2}$, $d_{3/2}$, and $d_{5/2}$ neutron partial waves.
For the harmonic oscillator basis we included all single-particle states
spanned by $17$ major oscillator shells.
\begin{figure}[thbp]
	\begin{center}
		\includegraphics[width=0.45\textwidth, clip]{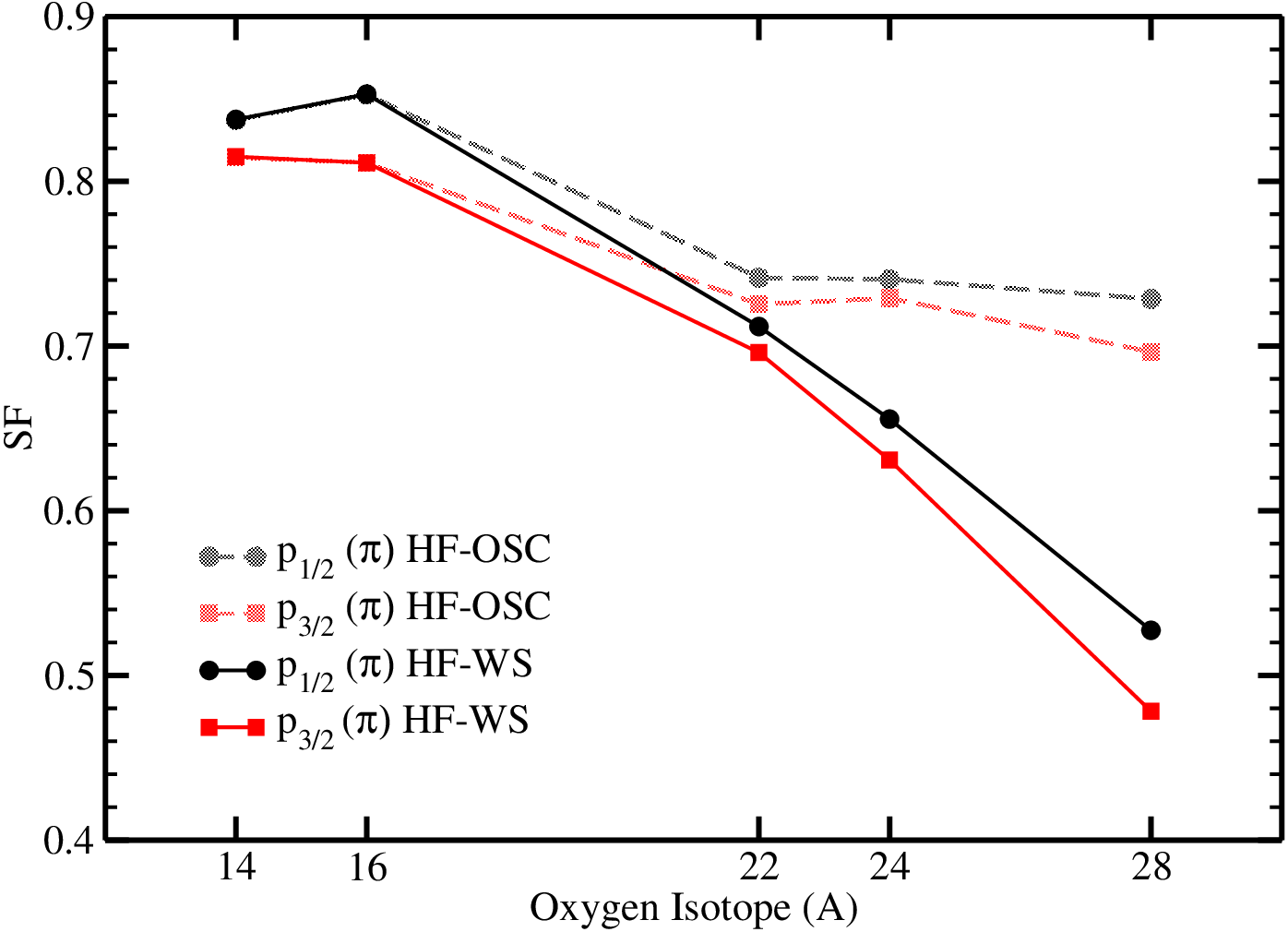}
	\end{center}
	\caption{\label{figSFvshw1}
	(Color online)
	Normalized spectroscopic factors for $p_{1/2}$ and $p_{3/2}$ proton removal from the 
        oxygen isotopes $^{14,16,22,24,28}$O. The continuum states
	included in the calculation (HF-WS) lead to a dramatic
        quenching of the spectroscopic factors 
        as the neutron dripline is approached. For comparison, we show calculations
        of spectroscopic factors using a HF basis built
        entirely from harmonic oscillator basis functions (HF-OSC). 
      }
\end{figure}

Figure \ref{figSFvshw1} shows the calculated SFs for
removing a proton in the $p_{1/2}$ and $p_{3/2}$ partial waves of
$^{14,16,22,24,28}$O. We compare our calculations of the SFs 
to calculations using an HF basis built entirely from
harmonic oscillator basis functions (HF-OSC, dashed lines).
The results are obtained with an harmonic oscillator energy 
$\hbar\omega = 30$ MeV.
Our calculations of the SFs depend weakly on the harmonic oscillator 
frequency, see for example Ref.~\cite{Jensen2010}.
The  $p_{1/2}$ and $p_{3/2}$  proton orbitals are close to the Fermi
level. In a traditional shell-model picture we would therefore expect
SFs close to unity for such states.
However, we find a significant quenching of the 
SFs due to the coupling-to-continuum degrees of
freedom. The calculations done with a HF-OSC basis show no significant 
quenching, and illustrate clearly the limitation of the harmonic 
oscillator basis representation of weakly-bound, neutron-rich nuclei. 
This observation agrees also nicely with the analysis of Michel 
{\em et al.} \cite{nicolas2007a}. There, the  authors
demonstrate that the energy dependence of SFs due to an opening of a
reaction channel can only be described properly in shell-model
calculations if correlations involving scattering states are treated
properly.  

In our calculations the closed-shell oxygen isotopes $^{14,16,22,24,28}$O are
all bound with respect to neutron emission 
(for this particular N$^3$LO interaction with cutoff $\Lambda=500$ MeV). In
particular, we get $^{28}$O bound by $3.67$ MeV with respect to
one-neutron emission. However, starting from a N$^3$LO with a cutoff
$\Lambda=600$ MeV, we get $^{28}$O unbound with respect to
four-neutron emission and $^{24}$O, as seen in Ref.~\cite{hagen2009}. 
Clearly, three-nucleon forces are needed, to decide whether theory 
predicts a bound or unbound $^{28}$O.    
We also computed SFs for the proton removal from $^{14,16,22}$O using the 
$\Lambda=600$ MeV N$^3$LO interaction model, and found similar 
results as for the $\Lambda=500$ MeV N$^3$LO interaction model. 

To further understand the role of correlations beyond mean-field 
we computed the SF for $p_{1/2}$  proton removal from $^{24}$O using 
the N$^3$LO interaction evolved to a lower momentum cutoff $\lambda$ 
through similarity renormalization group methods \cite{scott2007}. 
We used three different approximations to $\ket{A}$ and $\ket{A-1}$ 
and considered three different cutoffs 
$\lambda = 3.2,3.4,3.6 \mathrm{fm}^{-1}$. First, in the crudest
approximation, using a mean-field HF solution for $\ket{A}$ and
$\ket{A-1}$, the SFs are by definition equal to unity. 
Secondly, we used a HF solution for $\ket{A}$ while $\ket{A-1}$
was approximated by one-hole and two-hole-one-particle 
excitations on the HF ground state $\ket{A}$, in
this case we observed about $15-20\%$ reduction in the SFs. Finally, 
our EOM-CC approach in Eq.(\ref{equSFdefinition}), gave a reduction of
$20-25 \% $ over the range of cutoffs considered. 
This clearly shows the importance of correlations beyond the mean-field. 
Changing the cutoff from  $\lambda = 3.2 \mathrm{fm}^{-1}$ to 
$\lambda = 3.6 \mathrm{fm}^{-1}$, the SF varied from 0.79 to 0.75, 
illustrating the role of short range correlations on the
quenching of the spectroscopic factors for proton removal in $^{24}$O.

\begin{figure}[thbp]
	\begin{center}
		\includegraphics[width=0.45\textwidth, clip]{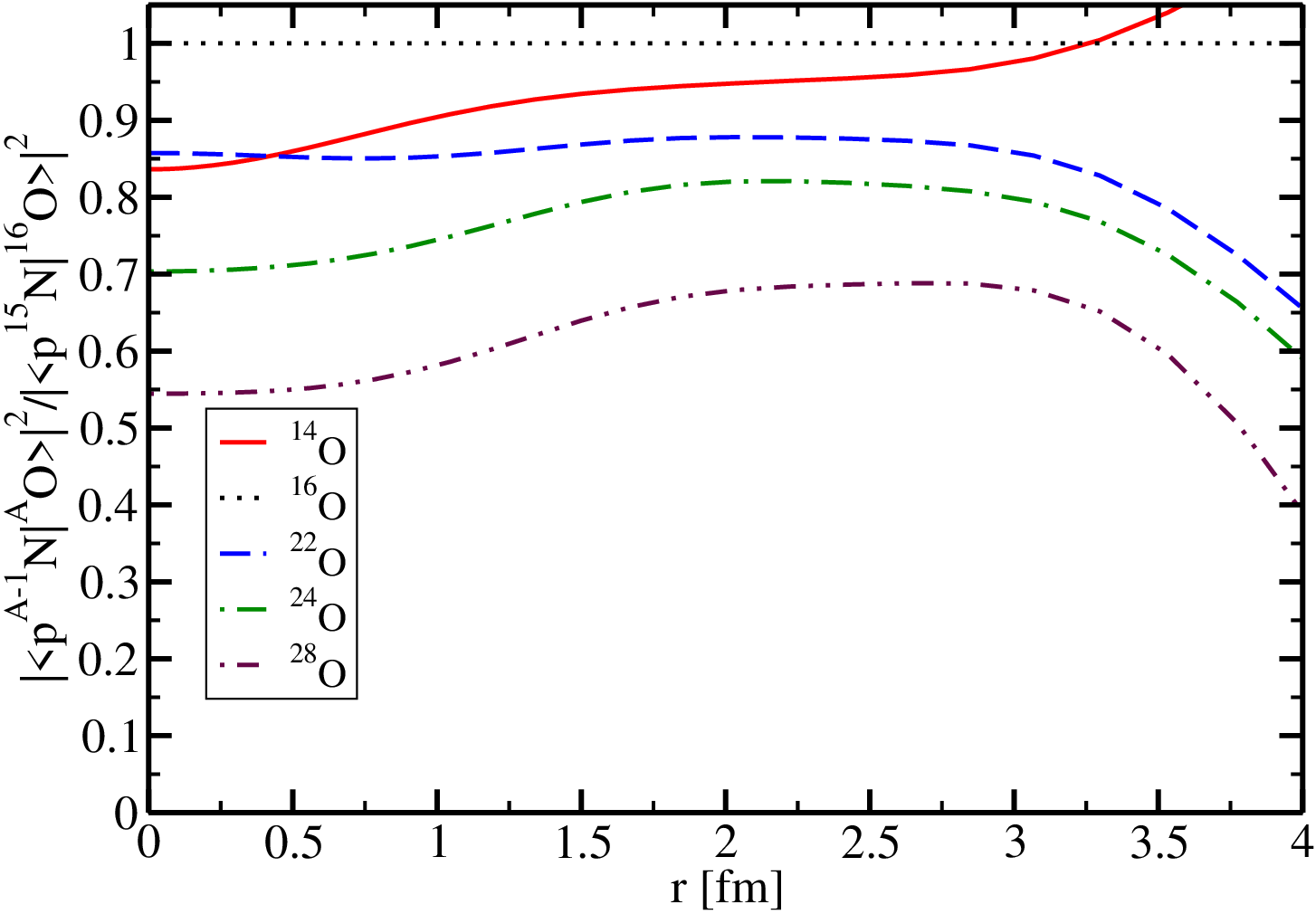}
	\end{center}
	\caption{\label{fig:overlap_p12}
	(Color online)
	Ratio of the radial overlap functions $ \langle ^{13}\mathrm{N}\vert
a_{lj}\vert^{14}\mathrm{O}\rangle, 
\langle ^{15}\mathrm{N}\vert a_{lj}\vert
^{16}\mathrm{O}\rangle, 
\langle ^{21}\mathrm{N}\vert a_{lj}\vert
^{22}\mathrm{O}\rangle, \langle ^{23}\mathrm{N}\vert a_{lj}\vert
^{24}\mathrm{O}\rangle,$ and $\langle ^{27}\mathrm{N}\vert a_{lj}\vert
^{28}\mathrm{O}\rangle $ for the $p_{1/2}$ single-particle state.}
\end{figure}

The SFs of proton removal in the oxygen isotopes are
determined by the squared norm of the overlap functions of Eq.~(\ref{equDefOverlap}).
In order to probe the sensitivity of the tail of the overlap functions 
as we move towards $^{28}$O, we compute the ratios of the absolute square of the
radial overlap functions to the $\vert \langle ^{15}\mathrm{N}\vert a_{lj}\vert
^{16}\mathrm{O}\rangle\vert^2 $ overlap function. These results are shown in
Fig.~\ref{fig:overlap_p12} for the 
$p_{1/2}$ proton state (the $p_{3/2}$ proton state shows a very similar pattern). 
A notable reduction of these norms towards more neutron-rich nuclei is
seen. The downward dip of the overlap ratios at larger radii comes 
from the fact the $p_{1/2}$ proton orbital become more and more bound 
as more neutrons are added to $^{16}$O. For $^{14}$O the $p_{1/2}$
proton is less bound with respect to $^{16}$O, resulting in a bend
upward. As the neutron dripline is approached, the one-neutron emission
thresholds for the oxygen isotopes and their neighboring
nitrogen isotopes are getting closer to the scattering
threshold. Clearly, the tail of the wave functions will play a more
important role as the outermost neutrons   
get closer to the scattering threshold. It is exactly this effect we 
observe in our calculations for the SFs for proton
removal. Using a HF basis of purely harmonic oscillator
wave functions, the density in the interior region of
the nucleus is overestimated, while the density is shifted towards the tail when using
a basis with correct asymptotic behavior.  
One should note that the nitrogen isotopes for a given neutron number are
more loosely bound than their corresponding oxygen isotones,
and this is the essential reason for the reduction.  
For $^{28}$O and $^{27}$N, no experimental 
values are available but if $^{28}$O exists it will be very loosely
bound and we may assume that  $^{27}$N is unbound.
 
\begin{figure}[thbp]
	\begin{center}
		\includegraphics[width=0.45\textwidth,clip]{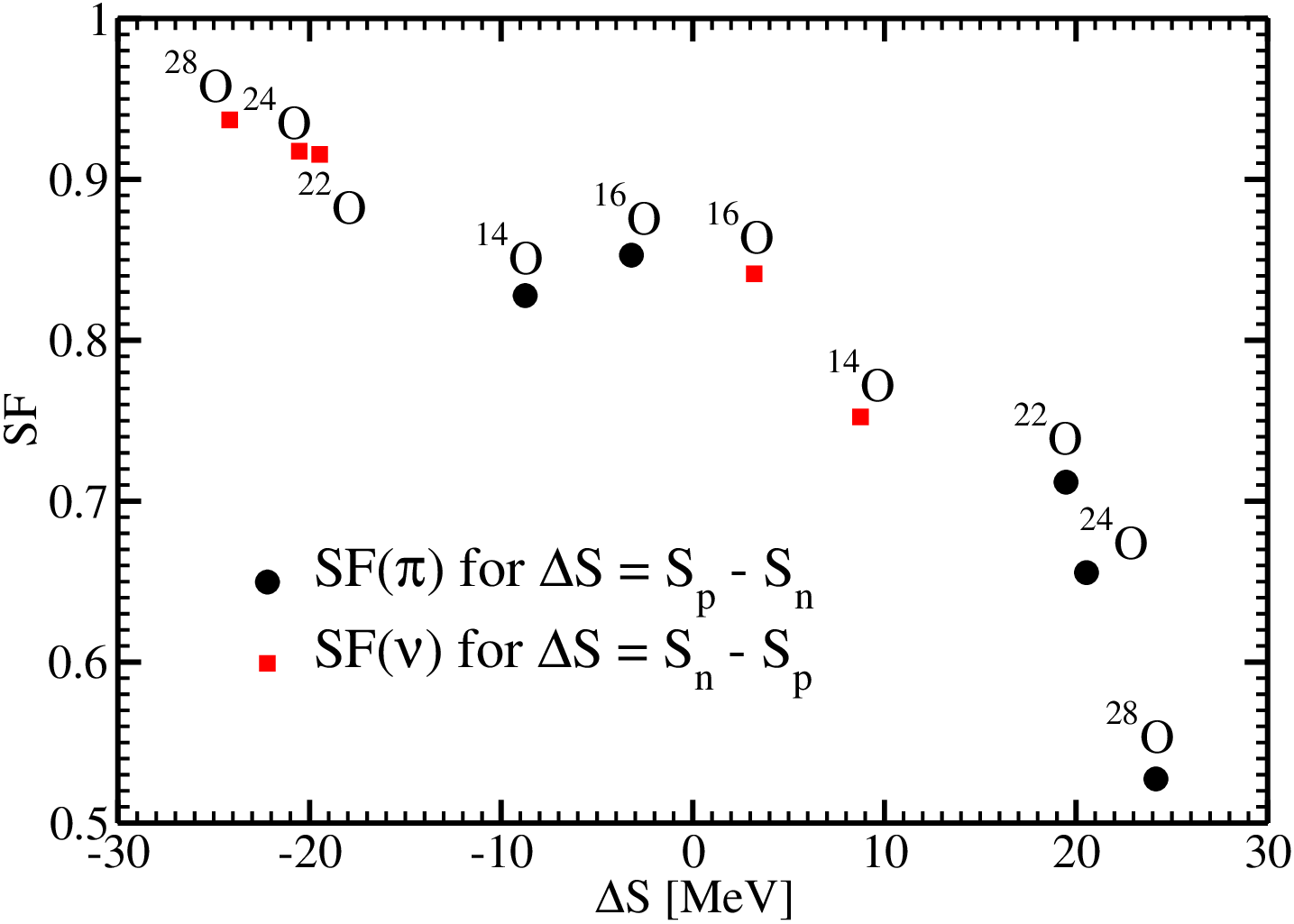} 
	\end{center}
	\caption{\label{fig:sfvsenergies}
		Plot of calculated SFs as functions of  
		the difference between the calculated neutron and proton separation energies.
		The results are for the single-particle states closest to the Fermi surface.
		For protons these are the $p_{1/2}$-states.}
\end{figure}

Finally, we show in Fig.~\ref{fig:sfvsenergies} the SFs of the proton and neutron
states closest to the Fermi surface (for protons the  
$p_{1/2}$-state), as a function of the difference between the computed 
proton and neutron separation energies. 
The results here agree excellently with similar interpretations made
in Refs.~\cite{alexandra2008, alexandra2004}. One sees clearly an
enhancement of correlations for the more strongly-bound, deficient  
nucleon species with increasing asymmetry.

In conclusion, 
we have found a large quenching for the spectroscopic
factors of the deeply-bound proton states near the Fermi surface in 
the neutron-rich oxygen isotopes.
This can be ascribed mainly to many-body correlations arising from a 
proper treatment of neutron scattering states. These results agree nicely with 
the mathematical analysis performed by Michel {\em et al} 
\cite{nicolas2007a}. This result for the oxygen
isotopes is similar to what has been inferred from neutron knockout
reaction cross sections for deeply-bound neutron states near
the Fermi surface in proton-rich $sd$-shell nuclei
\cite{alexandra2004,alexandra2008}. Clearly,  more
work is needed to confirm the connection; experiments
for proton knockout from oxygen should be undertaken and
many-body calculations for proton-rich, heavy nuclei need to be
carried out.

We thank Marek P{\l}oszajczak for useful comments.
This work was supported by the Office of Nuclear Physics, U.~S.~Department of Energy
(Oak Ridge National Laboratory); the University of
Washington under Contract No. DE-FC02-07ER41457.  
This research used computational
resources of the National Center for Computational Sciences and the Notur project in Norway.

\end{document}